\documentclass[aps,pra,superscriptaddress,twocolumn,floats,10pt,floatfix]{revtex4-1}
\usepackage{amsmath}
\usepackage{amsfonts}
\usepackage{amssymb}
\usepackage{graphicx}
\usepackage{color}
\usepackage[colorlinks=true,citecolor=blue,linkcolor=blue]{hyperref}
\usepackage{times}
\usepackage{amstext}
\usepackage{latexsym}

\begin{document}

\title{Relaxation time for monitoring the quantumness of an intense cavity field}
\author{D. Z. Rossatto}
\author{C. J. Villas-Boas}
\affiliation{Departamento de F\'{\i}sica, Universidade Federal de S\~{a}o Carlos, CEP 13565-905, S\~{a}o Carlos, SP, Brazil}

\begin{abstract}

Recently it was shown that the quantum behavior of an intense cavity field can be revealed by measuring the steady atomic correlations between two ideal atoms, which interact with the same leaking cavity mode. Considering a weak atom-field coupling regime and large average number of photons in the cavity mode ($\bar{n}$), one expects that a semiclassical theory could explain the whole dynamics of the system. However, this system presents the generation of correlations between the atoms, which is a signature of the quantumness of the cavity field, even in the limit of $\bar{n} \gg 1$ [Phys. Rev. Lett. \textbf{107}, 153601 (2011)]. Here, we extend this result by investigating the relaxation time for such a system. We have shown that the relaxation time of the system varies proportionally to $\bar{n}$ for a coherent driving, but it is inversely proportional to $\bar{n}$ for an incoherent pumping. Thus, the time required to observe the manifestation of the quantum aspects of a cavity field on the atomic correlations diverges as $\bar{n}$ tends to macroscopic values due to a coherent driving, while it goes to zero for incoherent pumping. For a coherent driving, we can also see that this system presents metastability, i.e., first the atomic system reaches a quasi-stationary state which lasts for a long time interval, but eventually it reaches the real steady state. We have also discussed the effects of small atomic decay. In this case, the steady correlations between the atoms disappear for long times, but the intense cavity field is still able to generate atomic correlations at intermediate times. Then, considering a real scenario, we would be able to monitor the quantumness of a cavity field in a certain time interval.

\end{abstract}

\maketitle

\section{Introduction}

The observation of quantum phenomena in the macroscopic world is commonly a hard task, as the quantum-to-classical transition (QCT) is expected in the limit of large excitation numbers and many-body systems \cite{bohr1976}. The origin for such a transition has been related to decoherence due to interaction with an environment \cite{schlosshauer2007}, unattainability of superposition of macroscopically distinguishable states \cite{ghirardi1986}, and imprecise measurements \cite{kofler2007,kim2014}. Furthermore, the QCT may depend on how a system is monitored, since the dynamics of a quantum system is usually disturbed by a meter \cite{Zurek2003}.

A two-level system is often employed to monitor the properties of cavity fields \cite{everit2009,langbein2013,houck2014,fink2010,optics1992}. For instance, in Refs. \cite{fink2010,optics1992} the QCT of a cavity mode was investigated by raising its average number of photons, $\bar{n}$, so that a classical behavior for the cavity transmission \cite{fink2010} and the normalized second order correlation function \cite{optics1992} were observed in the limit of the macroscopic field ($\bar{n}\gg 1 $), as expected.

Recently, we also investigated the QCT by raising $\bar{n}$ either coherently (coherent driving field) or incoherently (temperature)  \cite{rossatto2011}. We showed theoretically that the nonclassical behavior of a cavity field can be revealed even when $\bar{n}\gg 1 $, regime in which the statistical properties of the cavity field can be classically described \cite{fink2010,optics1992}. To do that, a pair of identical two-level atoms sufficiently far from each other, to be considered noninteracting systems, was employed to monitor the field behavior instead of a single one, i.e., the whole atomic system was used as a detector for the field behavior. 

A classical theory describing such a field is not able to explain the generation of any kind of correlations between the noninteracting atoms, since classical fields perform only local operations on them \cite{classical-field}. Thus, given the atomic system initially in a separable state, if the steady atomic state is (quantum or classically) correlated, the field is nonclassical. Although there is no steady entanglement, our previous results revealed that there are steady correlations between the atoms (classical correlations \cite{vedral2001}, quantum discord \cite{zurek2001} and, consequently, mutual information \cite{nielsen}) even when $\bar{n}\gg 1 $, i.e., the quantum character of the field is still there even for a macroscopic field \cite{rossatto2011}. As in \cite{everit2009}, we neglected the atomic decay since the atoms act as a meter to monitor the behavior of the cavity mode, i.e., an ideal detector was assumed.

Whereas long relaxation times are desired to avoid decoherence in quantum information processing \cite{nielsen}, short ones are also an object of interest when dissipative processes are engineered for preparing an specific steady state \cite{kraus2008,verstraete2009,prado2009,sorensen2011,li2012}. The knowledge of the system dynamics provides us the time scales in which the system evolves unitarily and then reaches its steady state. In this sense, because system behavior has only been analyzed in the stationary regime in Ref.~\cite{rossatto2011}, here we extend such results by investigating the long-times dynamics of that driven-dissipative open quantum system, showing that the interplay between driving agents, dissipation and internal interaction can substantially modify the relaxation times of the system. Moreover, we also show how the atomic spontaneous emission (real detector) affects the previous results. Therefore, our theoretical study presented here provides valuable and relevant information to perform an experimental demonstration of the quantumness of an intense cavity field, as well as to implement theoretical proposals based on the same driven-dissipative open quantum system, such as the generation of Werner-like stationary states \cite{jakobczyk2005,agarwal2006,jakobczyk2009,jin2013} and subradiant entangled states \cite{rossatto2013}.

In order to investigate the relaxation time of the system, we study the scaling of the spectral gap of the Liouvillian \cite{prosen2008,cai2013,jiang2014,znidaric2015} with the driving agents, dissipation and internal interaction. We show that, while the relaxation time of the system varies proportionally to $\bar{n}$ for a coherent driving, it varies inversely proportionally to $\bar{n}$ for an incoherent pumping (i.e., by increasing the reservoir temperature). Thus, the time required to observe the manifestation of the quantum aspects of a cavity field on the atomic correlations diverges as $\bar{n}$ tends to macroscopic values due to coherent driving. On the other hand, this time goes to zero for incoherent pumping.

The paper is organized as follows. Section II outlines the model describing the dynamics of the driven-dissipative open quantum system. In Sec.\ III, the concepts of spectral gap and relaxation time are introduced via the Liouvillian spectrum. Section IV includes our investigation of the spectral gap involving an intense cavity field generated either coherently or incoherently. The effects of small atomic spontaneous emission are presented in Sec.\ V. Finally, Sec.\ VI covers the summary of the results.

\section{Model}

The system comprises a driven cavity mode interacting resonantly with a pair of identical noninteracting two-level atoms ($\left\vert \text{g}\right\rangle $ = \ ground state, $\left\vert \text{e}\right\rangle $ = excited state), with coupling strength $g_{0}$. The total Hamiltonian which describes this system in the interaction picture is ($\hslash =1$) 
\begin{equation}
H=H_{\text{TC}}+H_{d},  \label{Htotal}
\end{equation}
where $H_{\text{TC}}=g_{0}(aS_{+}+a^{\dag }S_{-}) $ is the interaction part of the Tavis--Cummings Hamiltonian \cite{Tavis}, and $H_{d}=i\varepsilon (a^{\dag }-a) $ describes a driving field pumping the cavity mode with strength $\varepsilon$. The operators $S_{\pm }={\sum\nolimits_{j=1}^{2}}\sigma _{\pm }^{j}$ are the collective spin operators with $\sigma_{\pm}^{j}=(\sigma _{x}^{j}\pm i\sigma _{y}^{j}) /2$, in which $\sigma_{x,y,z}^{j}$ are the Pauli operators for each atom. The annihilation (creation) operator of the cavity field is indicated by $a$ $\left(a^{\dagger }\right) $.

We assume that each part of the system is locally coupled to a bosonic thermal bath, which is valid when the atom-field coupling $g_{0}$ is much smaller than the bare frequencies of the cavity mode and the atomic transition. Under the Born-Markov approximation, the dynamics is given by the following Lindblad master equation \cite{Breuer} 
\begin{eqnarray}
\dot{\rho} &=&-i\left[ H,\rho \right] +\kappa (n_{\text{th}}+1) 
\mathcal{D}[a] \rho +\kappa n_{\text{th}}\mathcal{D}[a^{\dag }] \rho \notag \\
&&+\frac{\gamma }{2}(n_{\text{th}}+1) {\sum\nolimits_{j}}%
\mathcal{D}[\sigma _{-}^{j}] \rho +\frac{\gamma }{2}n_{\text{th}}%
{\sum\nolimits_{j}}\mathcal{D}[\sigma _{+}^{j}] \rho ,    \label{eqmestraint}
\end{eqnarray}%
in which $n_{\text{th}}$ is the average number of thermal photons, $\kappa $ and $\gamma $  are the dissipation rates of the cavity mode and the atoms, respectively, and $\mathcal{D}[\mathcal{O}] \rho =(2\mathcal{O}\rho \mathcal{O}^{\dag }-\mathcal{O}^{\dag }\mathcal{O}\rho -\rho \mathcal{O}^{\dag }\mathcal{O}) $. 

By solving this equation it is possible to observe the time the system takes to reach the steady state, for instance. This relaxation time of the system can also be obtained by analyzing the Liouvillian spectrum: the eigenvalue of the Liouvillian with the smallest nonzero absolute real part determines the relaxation time, as we will discuss in the next section. The calculation of this spectrum for the general case is possible only numerically, allowing analytical solution in some particular cases. Fortunately, one of these cases is for large $\bar{n}$ and small atom-field couplings ($g_{0} \ll \kappa$), which is our main case of interest here.

\section{Liouvillian spectrum}

Equation (\ref{eqmestraint}) can be rewritten as
\begin{equation}
\dot{\rho}=\mathcal{L}\rho ,  \label{eqcompact}
\end{equation}
in which $\mathcal{L}$\ denotes the total Liouville operator (or Liouvillian), and whose solution can be obtained by solving the eigenvalue equation \cite{englert2003}
\begin{equation}
\mathcal{L}\rho _{\lambda }=\lambda \rho _{\lambda }.  \label{eveq}
\end{equation}
With the set of eigenvalues \{$\lambda $\} and eigenoperators \{$\rho _{\lambda }$\}, the state of the system is known for any time, namely,
\begin{equation}
\rho \left( t\right) =\sum\nolimits_{\lambda }c_{\lambda }e^{\lambda t}\rho
_{\lambda },  \label{rhot}
\end{equation}
in which \{$c_{\lambda }$\} denotes the coefficients of the decomposition of the initial state into the eigenoperator basis, $\rho \left( 0\right)=\sum\nolimits_{\lambda}c_{\lambda }\rho _{\lambda }$ \cite{englert2003}. 

Whereas the steady state, $\rho_{\text{ss}} \equiv \rho(t\to\infty)$, is related to the eigenoperators whose eigenvalues vanish, the real parts of the nonzero eigenvalues, which are negative in general $[\text{Re}{( \lambda)} < 0]$, give the relaxation rates of the system dynamics \cite{prosen2008,cai2013,jiang2014,znidaric2015}. In particular, the slowest nonzero rate is called the \textit{spectral gap} of the Liouvillian 
\begin{equation}
\Delta \equiv-\min_{\lambda \neq 0}\{\text{Re}(\lambda)\},  \label{spectralgap}
\end{equation}%
whose inverse sets the longest time scale for relaxation toward the steady state \cite{prosen2008,cai2013,jiang2014,znidaric2015}, i.e., $\|\rho(t)-\rho_{\text{ss}}\| \sim e^{-t/\tau}$ with $\tau \equiv 1/\Delta$ and $\|X\|=\text{Tr}(\sqrt{X^{\dagger}X})$ \cite{macieszczak2015}. Hence, the system reaches the steady state when $t\gg\tau$.

\section{Spectral gap and relaxation time}

In this section, we examine the system relaxation time for two cases in which a pronounced quantum behavior for the cavity mode is not expected, but emerges. First, we consider a coherently driven cavity mode $(\varepsilon \ne 0)$ at zero temperature $(n_\text{th} = 0)$, with $\varepsilon \gg \kappa \gg g_{0}$. Then, we consider a cavity mode at high temperature, $n_\text{th} \gg 1$, with $\varepsilon = 0$. Following the idea of Refs. \cite{everit2009,rossatto2011}, we neglect the atomic dissipation, adopting the atomic system as an ideal detector. The effects of a small atomic decay are discussed in Sec. V.

\subsection{Intense coherent field}

When $\bar{n}$ is controlled by a coherent driving field and the environment is at zero temperature, Eq. \eqref{eqmestraint} reduces to
\begin{equation}
\dot{\rho}=-i\left[ H,\rho \right] +\kappa \mathcal{D}\left[ a\right] \rho .
\label{emcohtotal}
\end{equation}

As the atomic system is our detector, the relaxation time for monitoring the quantumness of an intense cavity field is strictly related to the spectral gap of the reduced dynamics of the atomic system. Indeed, we will see that the spectral gap of the whole system is exactly the spectral gap of the atomic system.

To derive the reduced dynamics of the atomic system, we make use of three unitary transformations. The first one is exactly the displacement operator, $\mathcal{U}_{1}=\exp [-\varepsilon /\kappa ( a^{\dag }-a)] $ $(\rho _{1}=\mathcal{U}_{1}\rho \mathcal{U}_{1}^{\dag }) $, so that
\begin{equation}
\dot{\rho}_{1}=-i[H_{1},\rho _{1}] +\kappa \mathcal{D}[a] \rho _{1},  \label{rho1pt}
\end{equation}
in which $H_{1} = H_{\text{TC}}+\Omega S_{x}$, with $\Omega = g_{0}\varepsilon/\kappa$ and $S_{x}=(S_{+}+S_{-})$. By defining $J_{z}=\sum_{j}(\left\vert +_{j}\right\rangle \left\langle+_{j}\right\vert -\left\vert -_{j}\right\rangle \left\langle -_{j}\right\vert)$ and $J_{+} = J_{-}^{\dag } = \sum_{j}\left\vert+_{j}\right\rangle \left\langle -_{j}\right\vert$, with $\left\vert \pm_{j}\right\rangle =\left( 1/\sqrt{2} \right) \left( \left\vert \text{g}%
_{j}\right\rangle \pm \left\vert \text{e}_{j}\right\rangle \right)$, $H_{1}$ can be rewritten as
\begin{eqnarray}
H_{1} &=&\Omega J_{z}+\frac{g_{0}}{2}J_{z}(a^{\dag }+a)  \notag
\\
&&+\frac{g_{0}}{2}( J_{+}a+J_{-}a^{\dag }) -\frac{g_{0}}{2}(J_{+}a^{\dag }+J_{-}a).  \label{H1}
\end{eqnarray}%
It is worth stressing that, in this displaced picture, the average number of photons scales as $\bar n_{\text{dis}} \sim (g_{0}/\kappa)^{2}$ for $\varepsilon/\kappa \gg 1$ \cite{solano2003}, that is, the dimension of the required Fock basis to correctly describe the field variables is drastically reduced, representing a substantial gain in the computational effort.

When $2\Omega \gg g_{0}/2\Rightarrow 4\varepsilon /\kappa \gg 1$, an intense intracavity field is generated. In this case, the reduced atomic dynamics can be obtained by means of methods for effective dynamics developed in Refs. \cite{klimov2000,klimov2002,klimov2003}. To do that, we apply two unitary transformations over Eq. (\ref{rho1pt}), $\mathcal{U}_{2}=\exp[(g_{0}/4\Omega)( X_{+}-X_{-})] $ and $\mathcal{U}_{3}=\exp[-( g_{0}/4\Omega)( Y_{+}-Y_{-})] $, with $X_{+}=X_{-}^{\dag }=J_{+}a$ and $Y_{+}=Y_{-}^{\dag}=J_{+}a^{\dag }$. Thus, given that $\rho _{2}=\mathcal{U}_{3}\mathcal{U}_{2}\rho _{1}\mathcal{U}_{2}^{\dag }\mathcal{U}_{3}^{\dag }$ and $J_{x}=J_{+}+J_{-}$, the dynamics up to second order in $g_{0}/4\Omega$ is given by
\begin{eqnarray}
\dot{\rho}_{2} &\approx &-i[H_{2},\rho _{2}] + \kappa \mathcal{D}[a]\rho _{2} + \kappa \left( \frac{g_{0}}{4\Omega }\right)^{2} \mathcal{D}[J_{x}]\rho _{2}  \notag \\
&&-\kappa \left( \frac{g_{0}}{4\Omega }\right) ^{2}(2J_{z}a^{\dag}\rho _{2}a^{\dag }-J_{z}a^{\dag 2}\rho _{2}-\rho _{2}J_{z}a^{\dag 2}) \notag \\
&&-\kappa \left( \frac{g_{0}}{4\Omega }\right) ^{2}(2a\rho_{2}J_{z}a-J_{z}a^{2}\rho _{2}-\rho _{2}J_{z}a^{2})  \notag \\
&&+\kappa \left( \frac{g_{0}}{4\Omega }\right) (2a\rho_{2}J_{x}-J_{x}a\rho _{2}-\rho _{2}J_{x}a)  \notag \\
&&+\kappa \left( \frac{g_{0}}{4\Omega }\right) (2J_{x}\rho _{2}a^{\dag}-a^{\dag }J_{x}\rho _{2}-\rho _{2}a^{\dag }J_{x}) ,  \label{rho2pt}
\end{eqnarray}
with
\begin{eqnarray}
H_{2} &= &\Omega J_{z}+\frac{g_{0}}{2}J_{z}\left( a+a^{\dag }\right) - \frac{g_{0}^{2}}{8\Omega }J_{z}(a-a^{\dag }) ^{2}  \notag \\
&&-\frac{g_{0}^{2}}{8\Omega }(J_{+}^{2}+J_{-}^{2}) +\frac{g_{0}^{2}}{2\Omega }J_{z}J_{x}  \notag \\
&&+\frac{g_{0}^{2}}{4\Omega }(Y_{+}+Y_{-}-X_{+}-X_{-})(a^{\dag }+a) .  \label{H2}
\end{eqnarray}

In a new interaction picture, with respect to $H_{2}^{0} = \Omega J_{z}$, we can neglect fast-oscillating terms in Eq. \eqref{rho2pt} via the rotating wave approximation, as long as $\Omega \gg \kappa$ and, as stated before, $2\Omega \gg g_{0}/2$. Therefore,
\begin{eqnarray}
\dot{\rho}_{2} &\approx &-i[H_{3},\rho _{2}] +\kappa \mathcal{D}[a]\rho_{2}  \notag \\
&&-\kappa \left( \frac{g_{0}}{4\Omega }\right) ^{2}(2J_{z}a^{\dag}\rho _{2}a^{\dag }-J_{z}a^{\dag 2}\rho _{2}-\rho _{2}J_{z}a^{\dag 2}) \notag \\
&&-\kappa \left( \frac{g_{0}}{4\Omega }\right) ^{2}(2a\rho_{2}J_{z}a-J_{z}a^{2}\rho _{2}-\rho _{2}J_{z}a^{2})  \notag \\
&&+\kappa \left( \frac{g_{0}}{4\Omega }\right) ^{2}(\mathcal{D}[J_{-}] \rho_{2}+\mathcal{D}[J_{+}]\rho_{2}), \label{rho3pt}
\end{eqnarray}
with 
\begin{equation}
H_{3}=\Omega J_{z}+\frac{g_{0}}{2}J_{z}(a+a^{\dag }) - \frac{g_{0}^{2}}{8\Omega }J_{z}(a-a^{\dag })^{2}.  \label{H3}
\end{equation}

Finally, considering the bad-cavity limit, $\kappa \gg g_{0} $, the field variables can be adiabatically eliminated \cite{zoller2004}, so that the reduced atomic dynamics (for $\kappa t \gg 1$), in the interaction picture with respect to $H_{2}^{0}$, is
\begin{eqnarray}
\dot{\rho}_{\text{eff}}^{\text{at}} & = &\kappa \left( \frac{\kappa }{4\varepsilon }\right)^{2} (\mathcal{D}[J_{-}]\rho_{\text{eff}}^{\text{at}} + \mathcal{D}[J_{+}]\rho_{\text{eff}}^{\text{at}})  \notag \\
&&+\kappa \left( \frac{g_{0}}{2\kappa }\right)^{2}\mathcal{D}[J_{z}]\rho_{\text{eff}}^{\text{at}}.  \label{em_at_eff}
\end{eqnarray}
Under these conditions, intense intracavity field and bad-cavity limit, in which a pronounced quantum behavior for the field would not be expected since $\bar n \sim (\varepsilon/\kappa)^2 \gg 1$ \cite{fink2010,optics1992,rossatto2011}, we observe from Eq. \eqref{em_at_eff} that the atomic dynamics is characterized by the competition of two dynamics, one parametrized by $\Gamma _{\varepsilon }=\kappa (\kappa/4\varepsilon)^{2}$ while the other by $\Gamma _{g_{0}}=\kappa (g_{0}/2\kappa)^{2}$.

Because eigenvalues are invariants under unitary transformations (similarity transformation) \cite{arfken2005}, the spectrum of the atomic Liouvillian can be directly obtained from Eq. \eqref{em_at_eff}, whose eigenvalues (algebraic multiplicity) are $\lambda _{0}=0$ $(2)$, $\lambda_{1}=-4\Gamma_{\varepsilon }$ $(3)$, $\lambda _{2}=-12\Gamma_{\varepsilon }$ $(1)$, $\lambda _{3}=-4\Gamma _{g_{0}}-2\Gamma _{\varepsilon }$ $(6)$, $\lambda _{4}=-4\Gamma _{g_{0}}-10\Gamma _{\varepsilon }$ $(2)$ and $\lambda _{5}=-4\left( 4\Gamma _{g_{0}}+\Gamma _{\varepsilon}\right) $ $(2)$. 

Hence, for this coherent driving case, the spectral gap is
\begin{equation}
\frac{\Delta _{c}}{\kappa }=4\frac{\Gamma _{\varepsilon }}{\kappa }=\left( \frac{2\varepsilon}{ \kappa}\right)^{-2},  \label{Deltacoerente}
\end{equation}
so that the longest relaxation time toward the steady state is
\begin{equation}
\kappa \tau _{c}=\left( \frac{2\varepsilon }{\kappa }\right) ^{2},
\label{taucoerente}
\end{equation}
that is, the larger $\varepsilon$ (the more intense the intracavity field), the longer the relaxation time toward the steady state. Then, we note that the observation of the quantumness of an intense field, via the generation of steady correlations between the atoms, becomes a hard task for extremely intense fields, since the spectral gap vanishes, which implies that a very long time is needed to reach the steady state. In this case, the atomic decay might start to affect the dynamics, as it leads the atoms to the ground state, masking the measurements due to imperfections of the detector.

The vanishing of the spectral gap can lead to a nonequilibrium phase transitions \cite{prosen2008, znidaric2011, kessler2012, honing2012, horstmann2013, bianchi2014} and can result in a nonexponential relaxation toward a steady state \cite{cai2013, medvedyeva2014}. In the latter case, another striking feature can also occur, a partial relaxation into long-lived metastable states, which decay to the true steady state at much longer time, if there is a splitting in the spectrum of the Liouvillian \cite{macieszczak2015}. In fact, this separation of time scales is present here as we discuss below.

\begin{figure}[t]
\includegraphics[trim = 0mm 10mm 0mm 0mm, clip, width=0.45\textwidth]{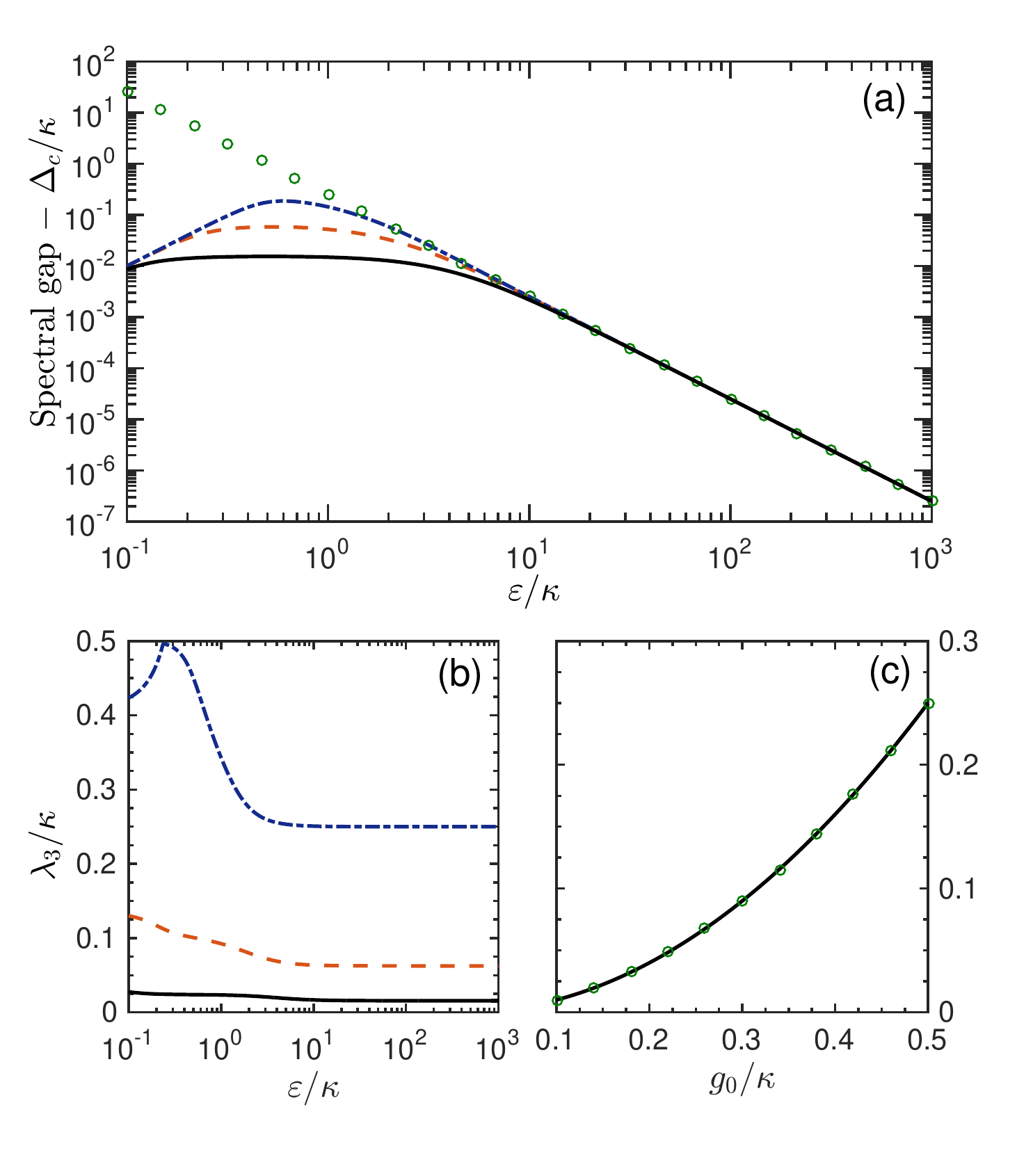}
\caption{(Color online) (a) Spectral gap vs $\varepsilon/\kappa$ for $g_0 = \kappa/8$ (black solid line), $g_0 = \kappa/4$ (red dashed line) and $g_0 = \kappa/2$ (blue dashed-dotted line) computed via the diagonalization of the exact Liouvillian of Eq. \eqref{rho1pt}. The green circles stands for Eq. \eqref{Deltacoerente}, which is in good agreement with the exact spectral gap for $\varepsilon/\kappa \gg 1$. (b) Second slowest nonzero rate vs. $\varepsilon/\kappa$ for the same $g_0$ of (a). (c) Second slowest nonzero rate vs. $g_0/\kappa$ for $\varepsilon = 100\kappa$. The green circles stands for the analytical result, $\lambda _{3}=-4\Gamma _{g_{0}}-2\Gamma _{\varepsilon }$, which is in good agreement with the exact result.}
\label{gapcoh}
\end{figure}

In Fig. \ref{gapcoh}(a), we observe, for $\varepsilon/\kappa \gg 1$ (regime in which the approximations are valid), a good agreement between the spectral gap computed exactly via the diagonalization of Eq. \eqref{rho1pt} [solid $(g_0 = \kappa/8)$, dashed $(g_0 = \kappa/4)$ and dashed-dotted lines $(g_0 = \kappa/2)$] and the analytical one given by Eq. \eqref{Deltacoerente} (green circles). The second slowest nonzero rate for the system relaxation, as a function of field intensity ($\varepsilon/\kappa$), is shown in Fig. \ref{gapcoh}(b) for the same values of $g_{0}$ used in Fig. \ref{gapcoh}(a), where we note that, for $\varepsilon/\kappa \gg 1$, it converges to a nonzero asymptotic value depending on $g_0$. This behavior indicates a separation of the time scales in the relaxation times of the Liouvillian, since the spectral gap vanishes. Considering $\varepsilon=100\kappa$,  Fig. \ref{gapcoh}(c) illustrates the second nonzero rate, as a function of $g_0/\kappa$, calculated via Eq. \eqref{rho1pt}, where the green circles stand for the analytical prediction, $\lambda _{3}=-4\Gamma _{g_{0}}-2\Gamma _{\varepsilon }$, which is also in a good agreement with the exact solution. Therefore, for an extremely intense field generated by a coherent driving field $(\varepsilon/\kappa \gg 1)$, while $\Gamma _{\varepsilon }$ rules the relaxation toward the steady state, $\Gamma _{g_{0}}$ determines the relaxation toward a metastable state.

The separation of time scales becomes clearer by analyzing the dynamics of observables. Since the generation of steady atomic correlations (quantum discord or mutual information, for instance) is a witness for the quantumness of an intense field \cite{rossatto2011}, we consider here the time evolution of the mutual information \cite{nielsen} between the atoms, which is defined as
\begin{equation}
\mathcal{I}(\rho _{\text{at}}) =S(\rho _{1}) +S(\rho_{2}) -S(\rho _{\text{at}}) ,
\label{mutual}
\end{equation}
in which $S(\rho) =\rho \log _{2}(\rho)$ is the von Neumann entropy, $\rho _{1(2)}$ is the reduced density operator of the atom $1(2)$, and $\rho _{\text{at}}$ is the density operator of the atomic system. 

\begin{figure}[t]
\includegraphics[trim = 0mm 15mm 0mm 15mm, clip, width=0.45\textwidth]{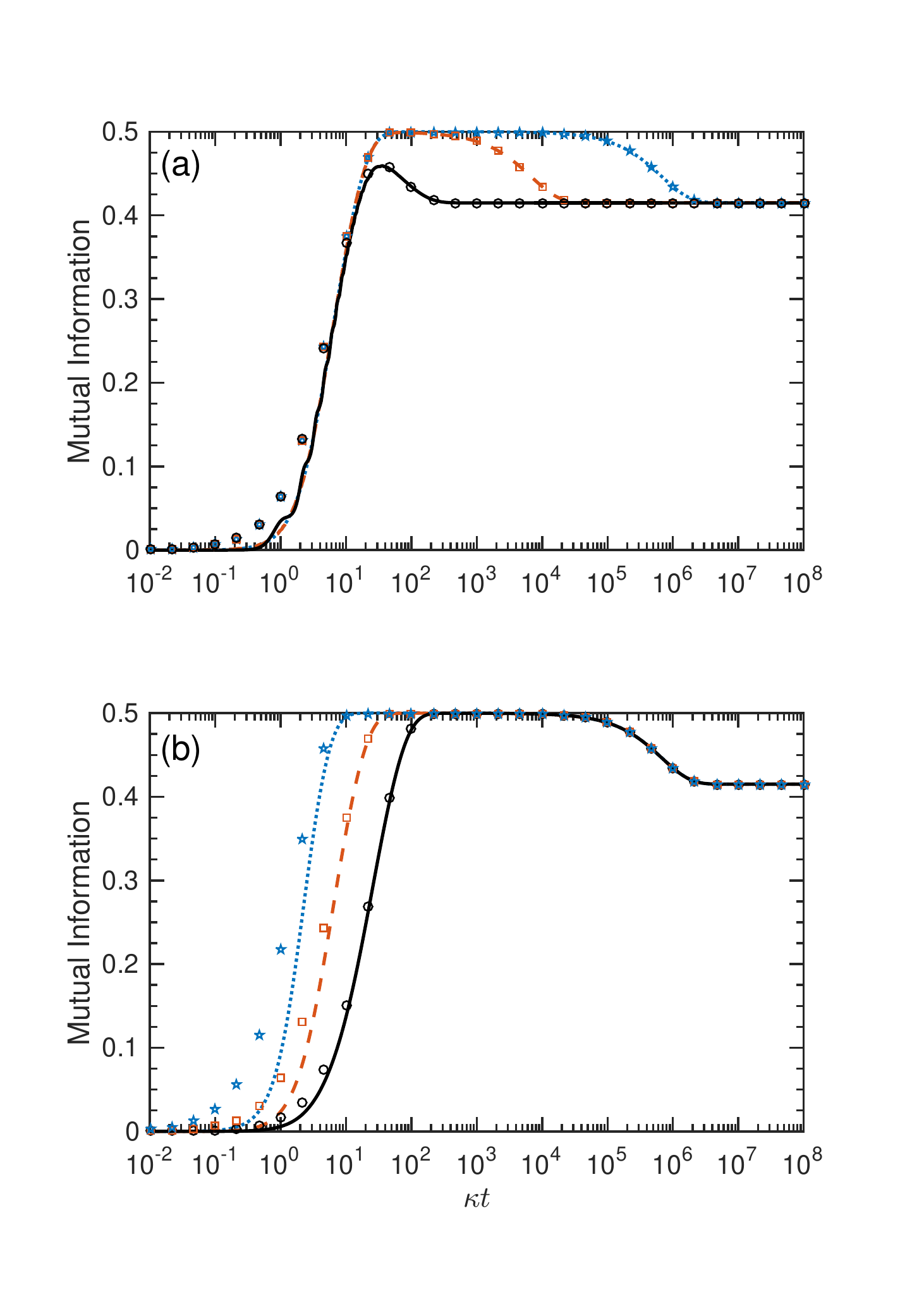}
\caption{(Color online) Time evolution of the mutual information between the atoms, making clear the separation of time scales yielding metastability. (a) $g_0=\kappa/4$ considering $\varepsilon=10\kappa$ (black solid line), $\varepsilon=100\kappa$ (red dashed line) and $\varepsilon=1000\kappa$ (blue dotted line). (b) $\varepsilon=1000\kappa$ considering $g_0=\kappa/8$ (black solid line), $g_0=\kappa/4$ and $g_0=\kappa/2$ (blue dotted line). While the lines are computed via Eq. \eqref{rho1pt}, the symbols are calculated using Eq. \eqref{em_at_eff}. We consider both atoms in the ground state and the cavity mode in the vacuum state initially.}
\label{corrcoh}
\end{figure}

In Fig. \ref{corrcoh}(a), we plot the mutual information for $g_0=\kappa/4$ (fixed) considering $\varepsilon=10\kappa$, $100\kappa$, and $1000\kappa$. Similarly, Fig. \ref{corrcoh}(b) stands for  $\varepsilon=1000\kappa$ (fixed) considering $g_0=\kappa/8$, $\kappa/4$, and $\kappa/2$. In addition to the good agreement between the exact solutions (lines), computed via Eq. \eqref{rho1pt}, and the approximate ones (symbols), computed using Eq. \eqref{em_at_eff}, it is straightforward to note the existence of metastability through the pronounced plateau in the dynamics of the mutual information, when there is a separation of time scales $(\tau_{c} \gg \tau'=1/\lambda_{3})$, so that the metastability occurs in the time window $\tau' \ll t \ll \tau_{c}$. In this case, the system seems to be stationary $(t \sim \tau')$, but eventually relaxes toward the true steady state $(t \sim \tau_{c})$ \cite{macieszczak2015}. In Fig. \ref{corrcoh}, we can also observe the influence of $g_{0}$ $(\varepsilon)$ on the system dynamics, the smaller $g_{0}$ (larger $\varepsilon$), the longer the relaxation time toward the metastable state (steady state).

\subsection{Intense incoherent field}

Let us consider now that the bath temperature controls $\bar{n}$ instead of a coherent driving field (i.e., $\varepsilon = 0$). In this case, Eq. \eqref{eqmestraint} reduces to
\begin{equation}
\dot{\rho}=-i[H_{\text{TC}},\rho] +\kappa (n_{\text{th}}+1) \mathcal{D}[a] \rho +\kappa n_{\text{th}}\mathcal{D}[a^{\dag }] \rho.  \label{emincohtotal}
\end{equation}
Assuming $\kappa \gg g_{0}$, the field variables can be adiabatically eliminated \cite{zoller2004}, so that the reduced atomic dynamics (for $\kappa t \gg 1$) in the interaction picture is given by
\begin{equation}
\dot{\rho}_{\text{at}}=\Gamma (n_{\text{th}}+1) \mathcal{D}[S_{-}] \rho _{\text{at}}+\Gamma n_{\text{th}}\mathcal{D}[S_{+}] \rho_{\text{at}}, \label{em_at_incoh}
\end{equation}%
with $\Gamma =\kappa (g_{0}/\kappa) ^{2}$. We note from Eq. \eqref{em_at_incoh} that the reduced dynamics effectively describes noninteracting atoms coupled to a common thermal bath.

The diagonalization of the Liouvillian of Eq. \eqref{em_at_incoh} provides the eigenvalues (algebraic multiplicity) $\lambda _{0}/\Gamma =0$ (2), $\lambda _{1}/\Gamma =-2n_{\text{th}}$ (2), $\lambda _{2}/\Gamma =-3(2n_{\text{th}}+1) +\sqrt{1+16n_{\text{th}}(n_{\text{th}}+1)}$ (2), $\lambda _{3}/\Gamma =-2(n_{\text{th}}+1)$ (2), $\lambda _{4}/\Gamma =-2(2n_{\text{th}}+1)$ (4), $\lambda _{5}/\Gamma =-4(2n_{\text{th}}+1) +4\sqrt{n_{\text{th}}(n_{\text{th}}+1) }$ (1), $\lambda _{6}/\Gamma =-3(2n_{\text{th}}+1) -\sqrt{1+16n_{\text{th}}(n_{\text{th}}+1)}$ (2) and $\lambda _{7}/\Gamma = -4(2n_{\text{th}}+1) -4\sqrt{n_{\text{th}}(n_{\text{th}}+1) }$ (1). Therefore, the spectral gap is
\begin{equation}
\frac{\Delta _{\text{inc}}}{\kappa }=2n_{\text{th}}\Gamma =2n_{\text{th}}\left( \frac{g_{0}}{\kappa }\right) ^{2},  \label{Deltainc}
\end{equation}
so that the longest relaxation time toward the steady state is
\begin{equation}
\kappa \tau _{\text{inc}}=\frac{1}{2n_{\text{th}}}\left( \frac{\kappa }{g_{0}}\right) ^{2}.  \label{tauinc}
\end{equation}

Here the spectral gap does not vanish for an intense intracavity field $(\bar n \sim n_{{th}} \gg 1)$. Indeed, contrary to the previous case (coherent driving), the larger the $n_{\text{th}}$ (i.e., the more intense the intracavity field), the shorter the relaxation time toward the steady state. Moreover, there is no splitting in the Liouvillian spectrum in this case, since all eigenvalues are proportional to $n_{\text{th}}$ for $n_{\text{th}} \gg 1$. Consequently, metastability due to separation of time scales in the relaxation times is not expected. Here, the spectral gap  depends on the ratio $g_{0}/\kappa$, which does not occur in the previous case. 

\begin{figure}[t]
\includegraphics[trim = 0mm 25mm 0mm 15mm, clip, width=0.45\textwidth]{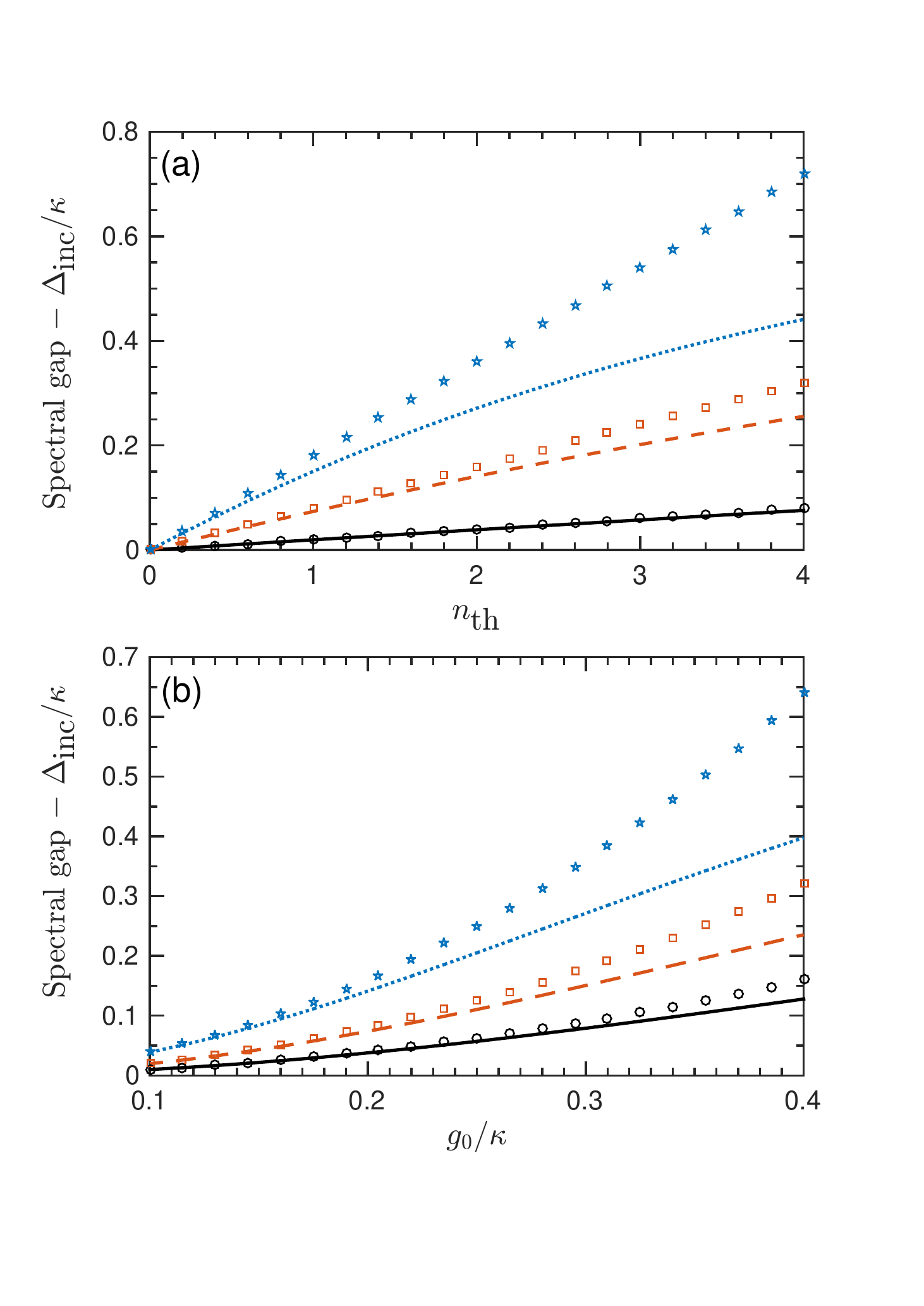}
\caption{(Color online) (a) Spectral gap vs $n_{\text{th}}$ for $g_{0} = 0.1\kappa$ (black solid line), $g_{0} = 0.2\kappa$ (red dashed line), and $g_{0} = 0.3\kappa$ (blue dotted line). (b) Spectral gap vs $g_{0}/\kappa$ for $n_{\text{th}} = 0.5$ (black solid line), $n_{\text{th}} = 1.0$ (red dashed line) and $n_{\text{th}} = 2.0$ (blue dotted line).}
\label{gapinc}
\end{figure}

The spectral gap of Eq. \eqref{emincohtotal} is displayed in Fig. \ref{gapinc} as a function of $n_{\text{th}}$ [Fig. \ref{gapinc}(a)] for $g_{0} = 0.1\kappa$, $0.2\kappa$, and $0.3\kappa$, and as a function of $g_{0}/\kappa$ [Fig. \ref{gapinc}(b)] for $n_{\text{th}} = 0.5$, $1.0$ and $2.0$. The symbols stand for the analytical solution given by Eq. \eqref{Deltainc}, which is in good agreement with the exact solution when $g_{0} \ll \kappa$, the regime in which the implemented approximations are valid. From this figure, we can note that the stronger the atom-field coupling $g_{0}$, the more pronounced the difference between the results predicted by the effective and the exact master equations. 

It is worth stressing that, in this case, as far as we know, we cannot perform an unitary transformation leading to a picture in which the required dimension of the Fock basis to correctly describe the field variables is reduced, as we carried out in the previous case. Therefore, this imposes a computational limitation in using Eq.  \eqref{emincohtotal} for very high values of $n_{\text{th}}$. However, even for small $n_{\text{th}}$, we observe a good agreement between the results predicted by the effective [Eq. \eqref{em_at_incoh}] and exact [Eq. \eqref{emincohtotal}] master equations.

\begin{figure}[t]
\includegraphics[trim = 0mm 5mm 0mm 5mm, clip, width=0.45\textwidth]{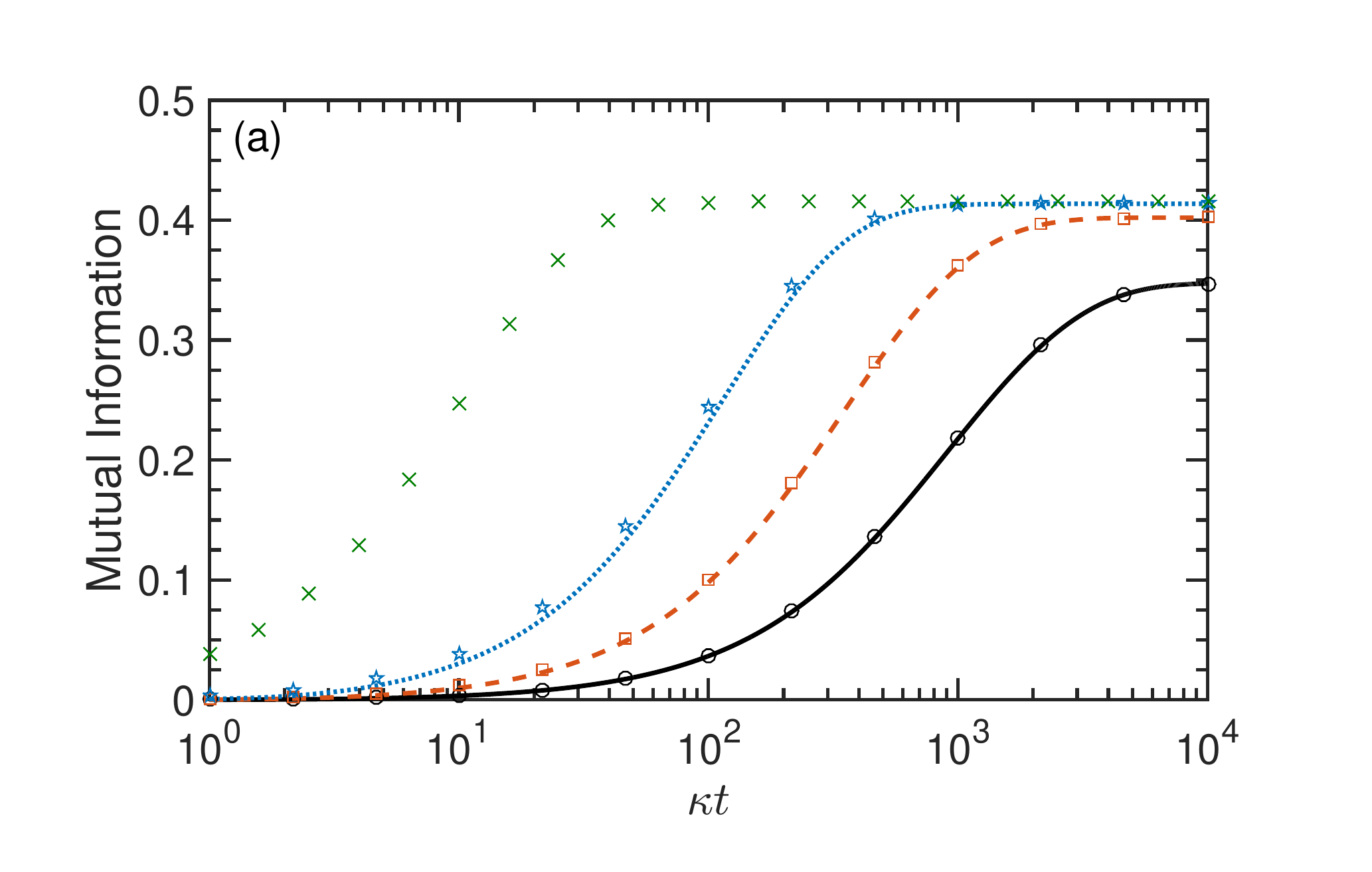}
\includegraphics[trim = 0mm 5mm 0mm 10mm, clip, width=0.45\textwidth]{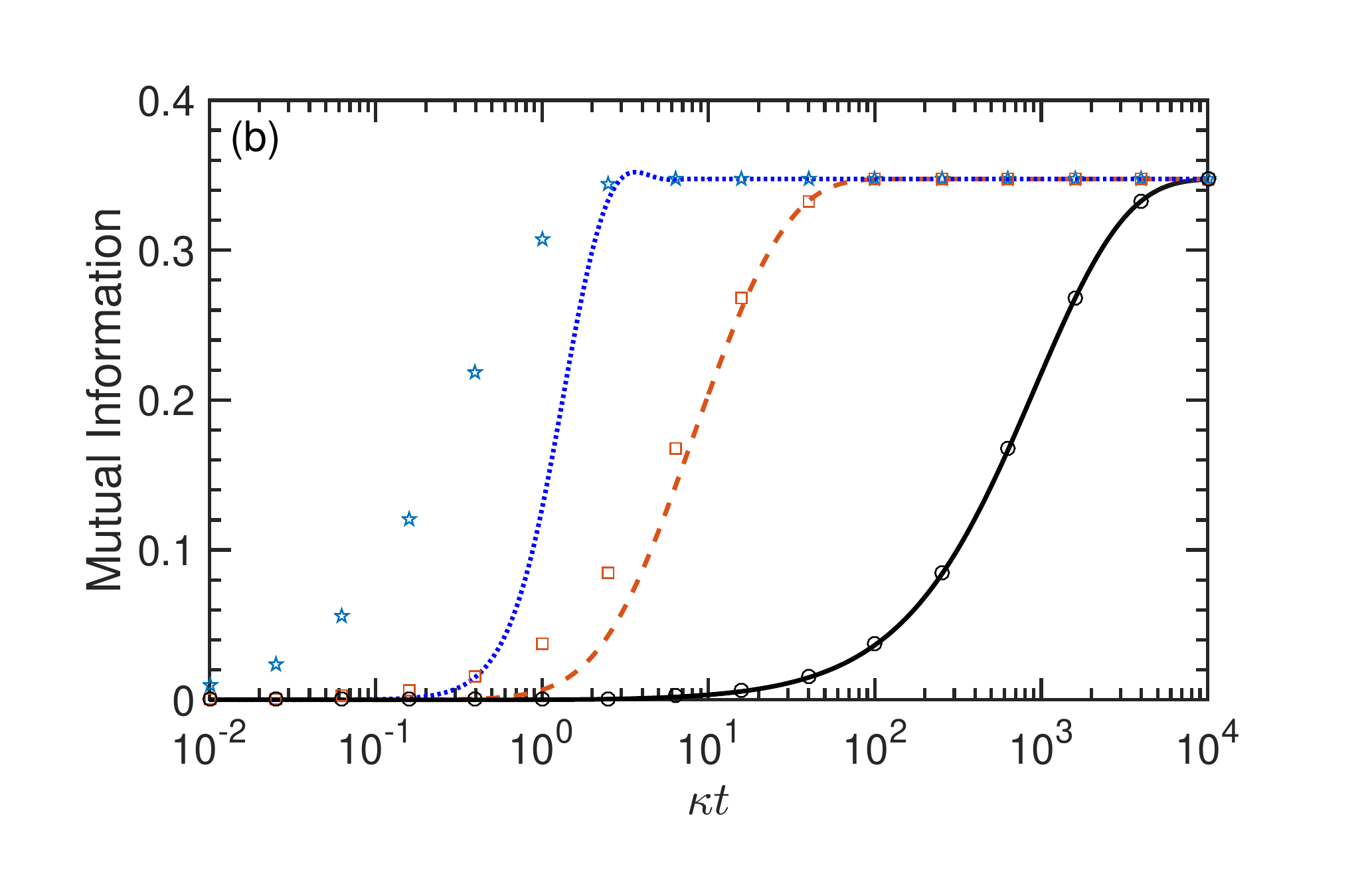}
\caption{(Color online) Time evolution of the mutual information between the atoms. (a) $g_0=\kappa/100$ considering $n_{\text{th}}=1$ (black solid line), $n_{\text{th}}=3$ (red dashed line), and $n_{\text{th}}=10$ (blue dotted line). (b) $n_{\text{th}}=1$ considering $g_0=\kappa/100$ (black solid line), $g_0=\kappa/10$ (red dashed line), and $g_0=\kappa/2$ (blue dotted line). While the lines are computed via Eq. \eqref{emincohtotal}, the symbols are calculated using Eq. \eqref{em_at_incoh}. In particular, the green crosses in (a) stand for $n_{\text{th}}=100$, showing that $n_{\text{th}}=10$ can already be characterized as $n_{\text{th}}\gg 1$, because of the saturation of the steady mutual information for $n_{\text{th}} \gtrsim 10$ \cite{rossatto2011}. We consider both atoms in the ground state and the cavity mode in the vacuum state initially.}
\label{corrincoh}
\end{figure}

In Fig. \ref{corrincoh}(a), we plot the time evolution of the mutual information calculated through both master equations for $g_{0} = 0.01 \kappa$ and considering $n_{\text{th}} = 1$, $3$, $10$, and $100$. Similarly, in Fig. \ref{corrincoh}(b), we have fixed $n_{\text{th}}=1$ considering $g_0=\kappa/100$, $\kappa/10$, and $\kappa/2$. The solution of the exact master equation was obtained by truncating the Hilbert space of the cavity mode and using the QuTiP master equation solver \cite{qutip}. From this figure, aside from the observation of the influence of $g_{0}$ and $n_{\text{th}}$ on the system dynamics, i.e., the smaller the $g_{0}$ or $n_{\text{th}}$, the longer the relaxation time toward the steady state, we clearly see that both master equations give the same result since we are within the validity of our approximations. For larger values of $g_{0}$ we can observe some discrepancy between the results predicted by the effective and exact master equations, but only concerning the beginning of the evolution. Both equations predict the same steady state even for stronger values of $g_{0}$, since they present the same eigenoperator with null eigenvalue for the Liouvillian. Moreover, it is worth noting the absence of metastability in this case.

Therefore, from an experimental point of view, the observation of the quantum behavior of an intense intracavity field by monitoring steady atomic correlations seems to be more feasible when the average number of intracavity photons is controlled incoherently, since the relaxation time decreases with the temperature, while it increases with the intensity of a coherent driving field.

\section{Real detector case}  

Here, the atomic system works out as a detector which monitors the cavity field behavior via the observation of steady correlations between the atoms. However, such correlations are highly sensitive to dissipative processes acting directly on the atoms \cite{werlang2009}, so that the absence of such correlations in the steady state might be due to the dissipative processes instead of the classicality of the field, masking the results. Consequently, considering a \textit{real detector} [$\gamma \neq 0$ in Eq. \eqref{eqmestraint}], our scheme is valid in a time window in which the atomic decay has not substantial influence on the atomic correlations yet. Moreover, it is worth stressing that all approximations performed in the previous section are valid only when $\gamma$ is much smaller than all other parameters in Eq. \eqref{eqmestraint}.

\begin{figure}[t]
\includegraphics[trim = 0mm 5mm 0mm 5mm, clip, width=0.45\textwidth]{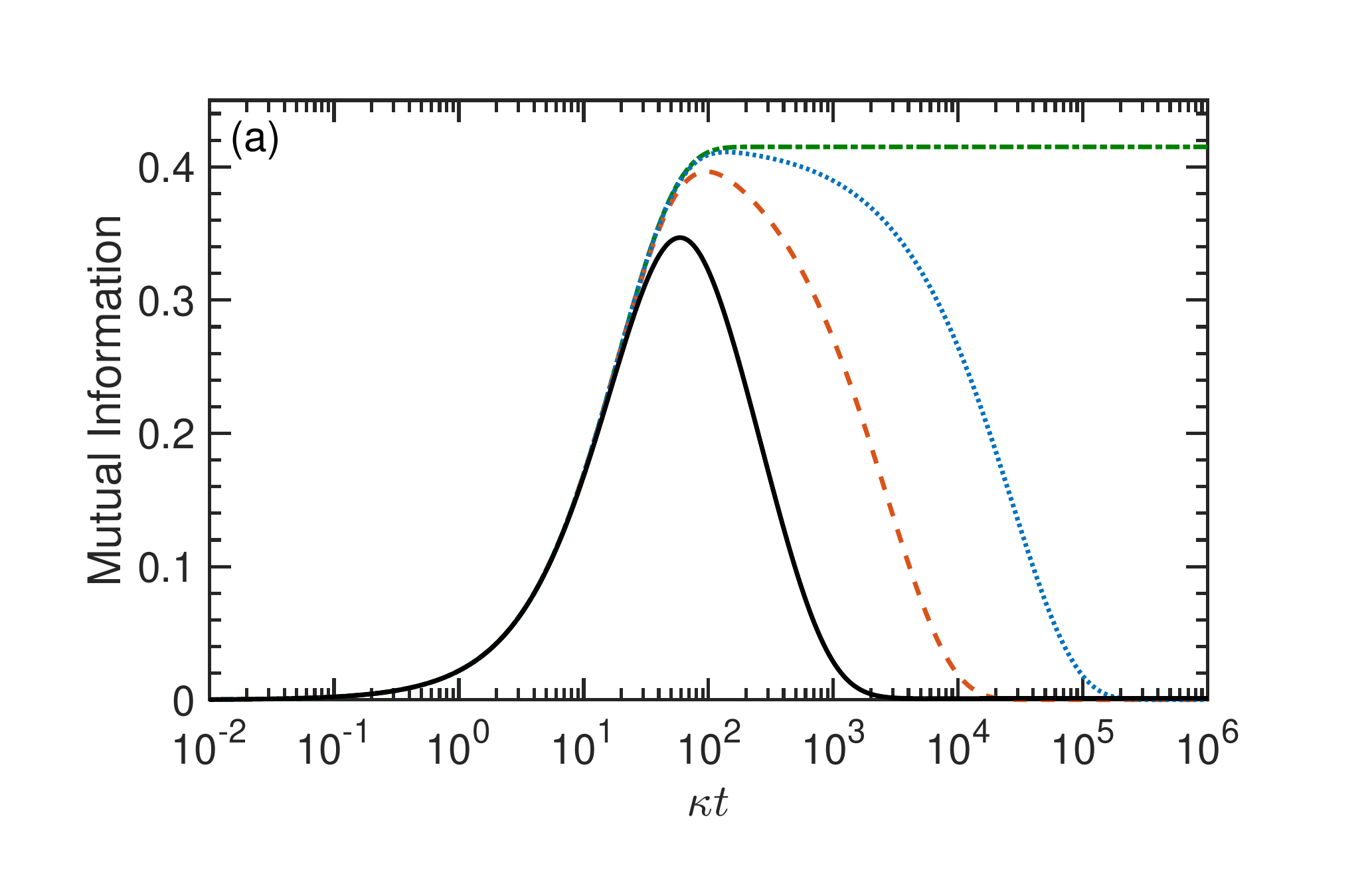}
\includegraphics[trim = 0mm 5mm 0mm 10mm, clip, width=0.45\textwidth]{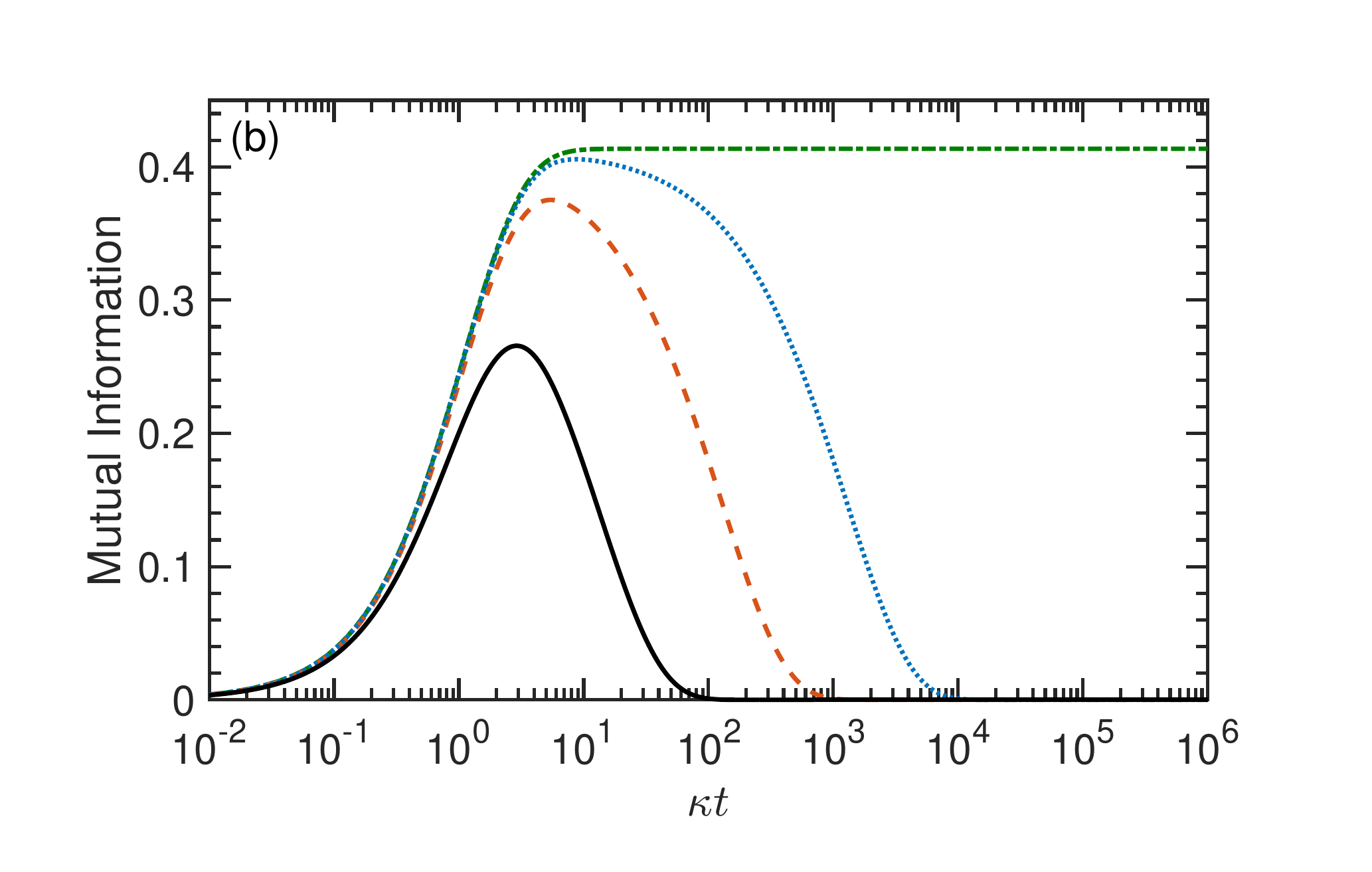}
\caption{(Color online) Time evolution of the mutual information considering different atomic decays: $\gamma =10^{-3}\kappa$ (black solid line), $\gamma =10^{-4}\kappa$ (red dashed line), $\gamma =10^{-5}\kappa $ (blue dotted line), and $\gamma = 0 $ (green dashed-dotted line), for the (a) coherent and (b) incoherent cases. We consider $g_{0}=0.1\kappa ,$ $\varepsilon =\sqrt{10}\kappa $ ($\bar{n}\sim 10$) in (a) and $n_{\text{th}}=10$ in (b). In both cases the atoms are initially in the ground state and the cavity mode in the vacuum.}
\label{Comp}
\end{figure}

Figure \ref{Comp} illustrates results similar to Figs. \ref{corrcoh} and \ref{corrincoh} taking into account the influence of small atomic decay for both coherent [Fig. \ref{Comp}(a)] and incoherent [Fig. \ref{Comp}(b)] cases, considering $g_{0}=0.1\kappa$. In both cases, the atoms are initially in the ground state and the cavity mode in the vacuum. The temperature of the reservoir and the strength of the driving coherent field were fixed, so that the average number of photons in the cavity mode was $\bar{n}\sim 10$. We can observe that it is always possible to see the generation of mutual information, a signature of the quantumness of the cavity mode. As it happened in the ideal detector situation ($\gamma = 0 $), the incoherent case is more feasible than the coherent one, since it takes much shorter times to generate atomic correlations. We also notice that the correlations disappear for sufficiently long times, making this system useful to reveal the quantum aspects of the cavity field only within a time window which depends on $\bar{n}$ of the cavity field and on $\gamma$.

\section{Summary}

We have investigated the relaxation time for two atoms interacting with intense cavity fields ($\bar{n} \gg 1$), generated either coherently (by continuous coherent driving on the cavity mode) or incoherently (via the interaction of the cavity mode with a thermal reservoir). We have also considered the weak atom-field coupling regime ($g_{0} \ll \kappa$), in which we usually expect that a semiclassical theory could explain the whole dynamics of the system. However, this system presents the generation of classical and quantum correlations between the atoms (except entanglement) even in the limit of $\bar{n} \gg 1$, which is a signature of the quantumness of the cavity field \cite{rossatto2011}.

We have shown that the time required to generate those correlations is proportional to the average number of photons in the cavity mode ($\bar{n}$) for the coherent case, but inversely proportional to $\bar{n}$ for the incoherent one. Therefore, an experimental observation of the quantum behavior of an intense intracavity field by monitoring steady atomic correlations seems to be more feasible when  $\bar{n}$ is controlled incoherently. When $\bar{n}$ is controlled coherently, we have also shown that our system presents metastability due to a splitting in the spectrum of the Liouvillian, i.e., first the atomic system reaches a long-lived metastable state, which decays to the real steady state at much longer time. We have also discussed the effects of small atomic decay. In this case, the steady correlations between the atoms disappear for long times, but the intense cavity field is still able to generate atomic correlations at intermediate times. Then, considering a real scenario, we would be able to monitor the quantumness of a cavity field in a certain time interval.

Finally, it is important to mention that the study presented here can be experimentally investigated with the current technology in the circuit QED scenario, for instance, as it was carried out in the experiment for a single artificial atom coupled to a cavity mode \cite{fink2010}, or following the work by J. Majer \textit{et al.} \cite{Majer2007} where two superconducting qubits were coupled to a resonator, exactly as required by our proposed scheme. Thus, we believe that our work contributes to the understanding of the quantum-to-classical transition of cavity fields, and provides the required time to experimentally perform the implementation of theoretical proposals based on the same driven-dissipative open quantum system, such as the generation of Werner-like stationary states \cite{jakobczyk2005,agarwal2006,jakobczyk2009,jin2013} and subradiant entangled states \cite{rossatto2013}.

\begin{acknowledgments} 

The authors gratefully acknowledge financial support from Brazilian agencies: Grants \#2013/23512-7 and \#2013/04162-5  S\~{a}o Paulo Research Foundation (FAPESP), Brazilian National Institute of Science and Technology for Quantum Information (INCT-IQ), and CNPq.

\end{acknowledgments}


\begin{thebibliography}{99}
\bibitem{classical-field} By classical field we mean a multimode field where the quantum annihilation/creation operators are replaced by complex time-dependent amplitudes, whose interaction with an atomic system is described by semiclassical differential equations (Maxwell-Bloch equations).
\end{thebibliography}
\end{document}